\documentclass[12pt]{article}
\usepackage{amsmath}
\def\a{\alpha}
\def\b{\beta}

\def\d{\delta}
\def\e{\epsilon}

\def\g{\gamma}
\def\p{\psi}

\def\n{\nu}

\def\s{\sigma}

\def\x{\xi}

\def\bp{\bar{\psi}}

\def\be{\begin{equation}}
\def\ee{\end{equation}}
\def\arr{\begin{array}{rll}}
\def\ea{\end{array}}
\def\bea{\begin{eqnarray}}
\def\eea{\end{eqnarray}}

\def\N2{$N{=}2$}

\def\>{\rangle}
\def\<{\langle}
\def\+{\dagger}
\def\={\ =\ }

\begin{document}
\renewcommand{\thefootnote}{\fnsymbol{footnote}}
\begin{titlepage}
\setcounter{page}{0}
\begin{flushright}
LMP-TPU--3/10  \\
\end{flushright}
\vskip 1cm
\begin{center}
{\LARGE\bf Conformal {} mechanics {} in }\\
\vskip 0.5cm
{\LARGE\bf Newton-Hooke spacetime  }\\
\vskip 1cm
$
\textrm{\Large Anton Galajinsky \ }
$
\vskip 0.7cm
{\it
Laboratory of Mathematical Physics, Tomsk Polytechnic University, \\
634050 Tomsk, Lenin Ave. 30, Russian Federation} \\
{E-mail: galajin@mph.phtd.tpu.ru}

\end{center}
\vskip 1cm
\begin{abstract} \noindent
Conformal many--body mechanics in Newton--Hooke spacetime is studied
within the framework of the Lagrangian formalism. Global symmetries and
Noether charges are given in a form convenient for analyzing the flat space limit.
$\mathcal{N}=2$ superconformal extension is built and
a new class on  $\mathcal{N}=2$ models related to
simple Lie algebras is presented. A decoupling similarity transformation
on $\mathcal{N}=2$ quantum mechanics in Newton--Hooke spacetime
is discussed.
\end{abstract}


\vskip 0.5cm


\end{titlepage}

\renewcommand{\thefootnote}{\arabic{footnote}}
\setcounter{footnote}0

\noindent
{\bf 1. Introduction}\\
\noindent

There are several reasons to be concerned about the conformal Newton-Hooke group and its dynamical realizations.
On the one hand, quantum mechanics of particles interacting via a conformal potential in arbitrary dimension
and confined in a harmonic trap has been extensively investigated over the last three decades
(see e.g. \cite{cm}--\cite{gm} and references therein). This study is motivated by the desire
to construct new exactly solvable many--body models in higher dimensions and explore novel correlations.
However, the fact that the underlying kinematical group is the conformal Newton-Hooke group does not seem to be widely appreciated.
On the other hand, recent proposals for a nonrelativistic version of the AdS/CFT
correspondence \cite{son,bala} stimulated a renewed interest in nonrelativistic conformal (super)algebras
\cite{say}--\cite{gl} (for related earlier studies see \cite{BDH}--\cite{hu}).
In this context it is natural to expect that many--body conformal mechanics in Newton-Hooke spacetime
will provide new insight into the correspondence.

Non--relativistic conformal algebras can be constructed in two different ways. One can either
minimally extend the Galilei algebra by generators of
dilatations and special conformal transformations  or consider non--relativistic
contractions of the relativistic conformal algebra
$so(d+1,2)$. In the former case one arrives at
the Schr\"odinger algebra \cite{nied,hagen}, while in the latter case the result is the conformal
Galilei algebra (see e.g.  \cite{olmo}--\cite{lsz1}). As
the conformal Galilei algebra implies vanishing mass,
the Schr\"odinger algebra has a better prospect
for quantum mechanical applications.

In the presence of a universal cosmological repulsion or attraction the Galilei algebra is replaced by the
Newton--Hooke algebra \cite{bac} (see also related works \cite{De}--\cite{Ti}).
The latter can be derived from the (anti) de Sitter algebra by a non-relativistic
contraction \cite{bac} and reproduces the Galilei algebra in the limit when the cosmological constant tends to zero.
Whereas in arbitrary dimension the Newton--Hooke algebra admits only one central charge,
in $(2+1)$--dimensions the second central charge is allowed which characterizes the exotic Newton--Hooke symmetry
\cite{gao,Olmo,Gom1,achp}. Extensions of the Newton--Hooke algebra by extra vector generators and their
dynamical realizations were discussed recently in \cite{luk2,Gom,tian}.
In the context of non--relativistic strings and branes
generalizations of the Newton--Hooke algebra were studied in \cite{sak,ggk,bgk,sy}.

Conformal extension of the Newton--Hooke algebra was discussed in \cite{olmo}.
However, its dynamical realization in many--body mechanics as
well as supersymmetric extension were unknown until quite recently \cite{gala1}.

The goal of this paper is to provide a systematic account of conformal many--body
mechanics in Newton--Hooke spacetime. We begin in Sect. 2 with a brief review of
conformal mechanics in flat space and the underlying Schr\"odinger group.
Sect. 3 is devoted to the Lagrangian formulation of conformal
mechanics in Newton--Hooke spacetime. The full list of global symmetries and the corresponding
Noether charges are given. $\mathcal{N}=2$ superconformal mechanics in flat space is discussed in Sect. 4 within the
Lagrangian formalism. Global symmetries which form the $\mathcal{N}=2$ Schr\"odinger superalgebra are derived.
Sect. 5 is focused on $\mathcal{N}=2$ mechanics in Newton--Hooke spacetime.
Negative and positive values of a cosmological constant are treated separately.
As shown below, $\mathcal{N}=2$ mechanics in Newton--Hooke spacetime is governed by a single scalar prepotential
which is translation and rotation invariant and obeys a first order linear partial differential equation.
In Sect. 6 we construct a novel class of $\mathcal{N}=2$ prepotentials which are related to simple Lie algebras.
Structure relations of $\mathcal{N}=2$ superconformal Newton--Hooke algebra are derived in Sect. 7 in the basis which
is convenient for studying the flat space limit. It is shown that in the case of a positive cosmological constant
the superalgebra is to be modified so as to take into account the spectrum of the energy operator.
In Sect. 8 we discuss a decoupling transformation which facilitates the analysis of
$\mathcal{N}=2$ quantum mechanics in Newton--Hooke spacetime.  We summarize our results
and discuss possible further developments in Sect. 9.

Throughout the work Greek letters are reserved for space indices $\a=1,\dots,d$. Latin letters
label identical particles $i=1,\dots,n$. Summation over repeated indices is understood.

\vspace{0.5cm}

\noindent
{\bf 2. Conformal mechanics in flat space}\\

\noindent

Conformal mechanics of $n$ identical particles in a flat space of arbitrary dimension is governed by the action functional
\be\label{act1}
S=\int dt \Big(\frac 12 \dot x^\a_i \dot x^\a_i-V(x) \Big).
\ee
For simplicity we set the mass of each particle to one.
It is assumed that the potential $V(x)$ obeys the set of linear partial differential equations
\be\label{restr}
\left(x_i^\a \partial_{\b i}-x_i^\b
\partial_{\a i}\right)V(x)=0, \qquad \sum_{i=1}^n \partial_{\a i} V(x)=0, \qquad x^\a_i \partial_{\a i} V(x)+2 V(x)=0,
\ee
where $\partial_{\a i}=\frac{\partial}{\partial x^\a_i}$. The first two constraints in (\ref{restr})
are needed to ensure the invariance of the action under space rotations, translations and Galilei boosts.
The last restriction provides conformal invariance.

The full symmetry group of the model is the Schr\"odinger group \cite{nied,hagen}. Apart from the
conventional time translations, space translations, space rotations and Galilei boosts, it involves
dilatations\footnote{Symmetry transformations which we consider in this work are of the form: $t'=t+\d t (t)$, $x'^{\a}_i (t')=x^\a_i(t)+\d x^\a_i(t,x(t))$.
If the action functional $S=\int dt \mathcal{L}(x,\dot x)$ holds invariant under the transformation up to a total derivative, i.e. $\d S=\int dt \Big( \frac{d F}{dt}\Big)$, then
the conserved quantity is derived from the expression
$\Big[\d x^\a_i \frac{\partial \mathcal{L}}{\partial \dot x^\a_i}-
\d t \Big( \dot x^\a_i \frac{\partial \mathcal{L}}{\partial \dot x^\a_i}-\mathcal{L} \Big)-F\Big]$ by discarding the parameter of the transformation.  }
\be
\d t=2t \lambda, \qquad \d x^\a_i=\lambda x^\a_i,
\ee
and special conformal transformations
\be
\d t=2 \s t^2, \qquad \d x^\a_i= 2\s t x^\a_i.
\ee
Denoting the corresponding conserved charges by $H$, $P^\a$, $M^{\a\b}$, $K^\a$, $D$ and $C$, respectively, one readily gets
\bea\label{Sch}
&&
H=\frac 12 \dot x^\a_i \dot x^\a_i+V(x), \qquad P^\a=\sum_{i=1}^n \dot x^\a_i, \qquad  M^{\a\b}=x^\a_i \dot x^\b_i-x^\b_i \dot x^\a_i,
\qquad K^\a=\sum_{i=1}^n x^\a_i-t P^\a,
\nonumber\\[2pt]
&&
D=-\frac 12 x^\a_i \dot x^\a_i +t H, \qquad C=\frac 12 x^\a_i x^\a_i+2t D-t^2 H.
\eea
Being rewritten in the Hamiltonian form, these functions form a representation of the Schr\"odinger algebra under the Poisson bracket
\cite{gala}.

\vspace{0.5cm}

\noindent
{\bf 3. Conformal mechanics in Newton--Hooke spacetime}\\

\noindent

In the presence of a universal cosmological repulsion or attraction the Schr\"odinger group is changed for the conformal
Newton-Hooke group \cite{olmo,gala1}. The former follows from the latter in the limit when the cosmological constant tends to zero.
In this Section we consider conformal many--body mechanics in Newton--Hooke spacetime, provide a complete account of its global
symmetries and construct the corresponding conserved charges. Negative and positive values of a cosmological constant are treated separately.

\vspace{0.5cm}

\noindent
3.1 {\it  Negative cosmological constant}\\

\noindent

In order to describe a universal cosmological attraction, one adds to the action functional (\ref{act1}) the extra harmonic
potential
\be\label{act2}
S=\int dt \Big(\frac 12 \dot x^\a_i \dot x^\a_i-V(x)-\frac{1}{2 R^2} x^\a_i x^\a_i \Big).
\ee
The corresponding coupling constant is proportional to the cosmological
constant \cite{bac}--\cite{Ti}. $R$ is the radius of a parent anti de Sitter space.

The new term in the action alters the form of all symmetries considered in the previous section, but for time translations and space rotations.
In particular,
one uncovers space translations
\be\label{T}
\d x^\a_i=a^\a \cos{(t/R)},
\ee
Newton--Hooke boosts
\be
\d x^\a_i=v^\a R \sin{(t/R)},
\ee
dilatations
\be
\d t= \lambda R \sin{(2t/R)}, \qquad  \d x^\a_i=\lambda x^\a_i \cos{(2t/R)},
\ee
and special conformal transformations
\be\label{S}
\d t= -\s R^2 (\cos{(2t/R)}-1), \qquad  \d x^\a_i=\s x^\a_i R \sin{(2t/R)},
\ee
provided the constraints (\ref{restr}) hold.
Application of the Noether theorem then yields the conserved charges
\bea\label{CH}
&&
H=\frac 12 \dot x^\a_i \dot x^\a_i+V(x)+\frac{1}{2 R^2} x^\a_i x^\a_i, \qquad P^\a=\Big(\sum_{i=1}^n \dot x^\a_i \Big) \cos{(t/R)}+
\frac{1}{R} \Big( \sum_{i=1}^n x^\a_i \Big) \sin{(t/R)},
\nonumber\\[2pt]
&&
M^{\a\b}=x^\a_i \dot x^\b_i-x^\b_i \dot x^\a_i, \qquad \qquad \quad
\qquad K^\a=\Big( \sum_{i=1}^n x^\a_i \Big) \cos{(t/R)}-\Big(\sum_{i=1}^n \dot x^\a_i \Big) R \sin{(t/R)},
\nonumber\\[2pt]
&&
D=-\frac 12 \Big( x^\a_i \dot x^\a_i \Big) \cos{(2t/R)} \ +\frac 12 H R \sin{(2t/R)}-\frac{1}{2 R} \Big( x^\a_i x^\a_i \Big) \sin{(2t/R)},
\nonumber\\[2pt]
&&
C=\frac 12 \Big( x^\a_i x^\a_i \Big) \cos{(2t/R)}-\frac 12 \Big( x^\a_i \dot x^\a_i \Big) R \sin{(2t/R)} -\frac 12 H R^2 (\cos{(2t/R)}-1).
\eea
We postpone calculation of the corresponding algebra to Sect. 7. For the moment
it suffices to observe that in the limit when the cosmological constant tends to zero (i.e. $R \to \infty$) both the symmetries and the
conserved charges transit smoothly to those in the previous section.

Note that the Newton-Hooke generators can also be obtained by exporting those
of the Galilei algebra using the conformal correspondence with the free case \cite{dhp}.
It should also be mentioned that some authors tend to consider the Schr\"odinger algebra and
the conformal Newton-Hooke algebra as one and the same thing written in different bases. For example, one
can rewrite the functions (\ref{CH}) within the framework of the Hamiltonian formalism
and set $t=0$ . This yields a representation of the conformal Newton-Hooke algebra under the Poisson bracket
in which
the oscillator potential entering the Hamiltonian coincides with the generator of special conformal
transformations up to a (dimensionful) constant. However, passing from the Schr\"odinger algebra to the
Newton-Hooke algebra in this manner would require a fundamental parameter of the dimension of length
which is not at our disposal in a flat space. This situation resembles the conformal particle in one dimension
studied in \cite{ddf}.

\vspace{0.5cm}

\noindent
3.2 {\it  Positive cosmological constant}\\

\noindent

The case of a universal cosmological repulsion is treated in a similar fashion. One flips the sign of the last term in (\ref{act2})
\be
S=\int dt \Big(\frac 12 \dot x^\a_i \dot x^\a_i-V(x)+\frac{1}{2 R^2} x^\a_i x^\a_i \Big),
\ee
and then changes trigonometric functions in (\ref{T})--(\ref{S}) by
hyperbolic ones in a proper way. This yields translations
\be
\d x^\a_i=a^\a \cosh{(t/R)},
\ee
boosts
\be
\d x^\a_i=v^\a R \sinh{(t/R)},
\ee
dilatations
\be
\d t= \lambda R \sinh{(2t/R)}, \qquad  \d x^\a_i=\lambda x^\a_i \cosh{(2t/R)},
\ee
and special conformal transformations
\be
\d t= \s R^2 (\cosh{(2t/R)}-1), \qquad  \d x^\a_i=\s x^\a_i R \sinh{(2t/R)}.
\ee
Time translations and space rotations remain intact.
As for the conserved charges, they take the following form
\bea
&&
H=\frac 12 \dot x^\a_i \dot x^\a_i+V(x)-\frac{1}{2 R^2} x^\a_i x^\a_i, \quad P^\a=\Big(\sum_{i=1}^n \dot x^\a_i \Big) \cosh{(t/R)}-
\frac{1}{R} \Big( \sum_{i=1}^n x^\a_i \Big) \sinh{(t/R)},
\nonumber\\[2pt]
&&
M^{\a\b}=x^\a_i \dot x^\b_i-x^\b_i \dot x^\a_i, \qquad \quad \quad
\qquad K^\a=\Big( \sum_{i=1}^n x^\a_i \Big) \cosh{(t/R)}-\Big(\sum_{i=1}^n \dot x^\a_i \Big) R \sinh{(t/R)},
\nonumber\\[2pt]
&&
D=-\frac 12 \Big( x^\a_i \dot x^\a_i \Big) \cosh{(2t/R)} \ +\frac 12 H R \sinh{(2t/R)}+\frac{1}{2 R} \Big( x^\a_i x^\a_i \Big) \sinh{(2t/R)},
\nonumber\\[2pt]
&&
C=\frac 12 \Big( x^\a_i x^\a_i \Big) \cosh{(2t/R)}-\frac 12 \Big( x^\a_i \dot x^\a_i \Big) R \sinh{(2t/R)} +\frac 12 H R^2 (\cosh{(2t/R)}-1).
\eea
Like in the previous subsection, it is readily verified that in the limit $R \to \infty$ both the symmetries and the
conserved charges go over to the corresponding expressions in Sect. 2.

\vspace{0.5cm}

\noindent
{\bf 4. $\mathcal{N}=2$ superconformal mechanics in flat space}\\

\noindent

We now discuss a supersymmetric extension of the conformal many--body mechanics in a flat space.
First of all, one has to decide on the number of supersymmetries to be realized.
For $\mathcal{N}=1$ one has a free theory. Beyond $\mathcal{N}=2$ it is problematic
to reconcile supersymmetric interactions and translation invariance (see e.g. \cite{bgl}) unless one introduces
spin degrees of freedom. By these reasons in what follows we focus on $\mathcal{N}=2$ supersymmetry.

In order to realize $\mathcal{N}=2$ supersymmetric extension of (\ref{act1}), one introduces fermionic degrees of freedom
$\p^\a_i$, $\bar\p^\a_i$ which are complex conjugates of each other. A single prepotential $U(x)$ is then used to generate both
the bosonic potential and the boson--fermion coupling
\be\label{act3}
S=\frac 12 \int dt \Big(\dot x^\a_i \dot x^\a_i-i \dot{\bar\p}^\a_i \p^\a_i+i\bar\p^\a_i {\dot\p}^\a_i
-\partial_{\a i} U(x) \partial_{\a i} U(x) +2 \partial_{\a i} \partial_{\b j} U(x) \p^\a_i \bar\p^\b_j \Big).
\ee
It is readily verified that this action holds invariant under transformations from the Schr\"odinger
group\footnote{The fermionic degrees of freedom transform as vectors under space rotations.
They are intact under all other transformations from the
Schr\"odinger group.} provided the prepotential $U(x)$ obeys the restrictions
\bea\label{Str2}
\left(x_i^\a \partial_{\b i}-x_i^\b
\partial_{\a i}\right)U(x)=0, \qquad
\sum_{i=1}^n \partial_{\a i} U(x)=0,\qquad
x_i^\a \partial_{\a i} U(x)=Z,
\eea
where $Z$ is a real constant. As above, the first two constraints are needed to provide the invariance of the action
under space translations, Galilei boosts and rotations, while the last one guarantees conformal invariance.
Note that in the supersymmetric case the bosonic potential factorizes as
\be\label{VB}
V_B(x)=\frac 12 \partial_{\a i} U(x) \partial_{\a i} U(x) .
\ee
That the latter is conformal is guaranteed by the rightmost equation in (\ref{Str2}).

Apart form the Schr\"odinger group, one reveals the supersymmetry
\bea
\d \p^\a_i=\Big(\dot x^\a_i -i\partial_{\a i} U(x) \Big) \bar\e, \qquad
\d \bar\p^\a_i=\Big(\dot x^\a_i +i\partial_{\a i} U(x) \Big) \e,
\qquad
\d x^\a_i =i\Big(\p^\a_i \e+\bar\p^\a_i \bar\e\Big),
\eea
and the superconformal symmetry
\bea
&&
\d \p^\a_i=-i t \Big(\dot x^\a_i-i\partial_{\a i} U(x) \Big) \bar\kappa+i x^\a_i \bar\kappa, \quad
\d \bar\p^\a_i=i t \Big(\dot x^\a_i+i\partial_{\a i} U(x) \Big) \kappa-i x^\a_i \kappa,
\nonumber\\[2pt]
&&
\d x^\a_i=-t\Big(\p^\a_i \kappa-\bar\p^\a_i \bar\kappa \Big).
\eea
Besides, in the sector of the fermionic variables one finds
the translations
\be
\d \p^\a_i=\x^\a, \qquad \d \bar\p^\a_i=\bar\x^\a,
\ee
and the $U(1)$--transformations
\be\label{u1}
\d \p^\a_i=i\n \p^\a_i, \qquad \d \bar\p^\a_i=-i\n \bar\p^\a_i.
\ee

The associated conserved charges are constructed with the use of the Noether
theorem\footnote{Symmetry transformations considered in this section are of the form: $t'=t+\d t (t)$,
$x'^{\a}_i (t')=x^\a_i(t)+\d x^\a_i(t,x(t),\p(t),\bp(t))$,
$\p'^{\a}_i (t')=\p^\a_i(t)+\d \p^\a_i(t,x(t),\p(t))$, $\bar\p'^{\a}_i (t')=\bar\p^\a_i(t)+\d \bar\p^\a_i(t,x(t),\bp(t))$.
If the action functional holds invariant up to a total derivative, $\d S=\int dt \Big( \frac{d F}{dt}\Big)$,
the conserved quantity is derived from the quantity $\Big[\d x^\a_i \frac{\partial \mathcal{L}}{\partial \dot x^\a_i}+
\d \p^\a_i \frac{\overrightarrow{\partial} \mathcal{L}}{\partial \dot \p^\a_i}+
\d \bar\p^\a_i \frac{\overrightarrow{\partial} \mathcal{L}}{\partial \dot \bp^\a_i}
-\d t \Big( \dot x^\a_i \frac{\partial \mathcal{L}}{\partial \dot x^\a_i}
+\dot \p^\a_i \frac{\overrightarrow{\partial} \mathcal{L}}{\partial \dot \p^\a_i}+
\dot \bp^\a_i \frac{\overrightarrow{\partial} \mathcal{L}}{\partial \dot \bp^\a_i}
-\mathcal{L} \Big)-F \Big]$ by discarding the parameter of the transformation.  }
\bea\label{ch1}
&&
H=\frac 12 \dot x^\a_i \dot x^\a_i+\frac 12 \partial_{\a i} U(x) \partial_{\a i} U(x)-\partial_{\a i} \partial_{\b j} U(x) \p^\a_i \bar\p^\b_j,
\qquad P^\a=\sum_{i=1}^n \dot x^\a_i ,
\nonumber\\[2pt]
&&
K^\a=\sum_{i=1}^n x^\a_i  -t P^\a, \qquad M^{\a\b}=x^\a_i \dot x^\b_i-x^\b_i \dot x^\a_i-i\Big(\p^\a_i \bar\p^\b_i-\p^\b_i \bar\p^\a_i \Big),
\nonumber\\[2pt]
&&
 \qquad \qquad \quad
\qquad
\nonumber\\[2pt]
&&
D=-\frac 12 x^\a_i \dot x^\a_i +t H, \qquad C=\frac 12 x^\a_i x^\a_i+2t D-t^2 H, \qquad J=\p^\a_i \bar\p^\a_i,
\nonumber\\[2pt]
&&
S=x^\a_i \p^\a_i-t Q, \quad \bar S=x^\a_i \bar \p^\a_i-t \bar Q, \quad L^\a=\sum_{i=1}^n \p^\a_i, \quad \bar L^\a=\sum_{i=1}^n \bar\p^\a_i,
\nonumber\\[2pt]
&&
Q=\Big(\dot x^\a_i +i \partial_{\a i} U(x) \Big) \p^\a_i, \quad
\bar Q=\Big(\dot x^\a_i -i \partial_{\a i} U(x) \Big) \bar\p^\a_i.
\eea
Here $J$ corresponds to the $U(1)$--transformation, $(S,\bar S)$ are related to the superconformal transformations,
$(L^\a,\bar L^\a)$ are linked to the translations in the fermonic sector, and $(Q,\bar Q)$ correspond to the supersymmetry transformations.
Being rewritten in the Hamiltonian form, these functions form a representation of the
$\mathcal{N}=2$ Schr\"odinger superalgebra under
the Poisson bracket \cite{gm}.

\vspace{0.5cm}

\noindent
{\bf 5. $\mathcal{N}=2$ superconformal mechanics in Newton--Hooke spacetime}\\

\noindent

Depending on the sign of a cosmological constant, the construction of $\mathcal{N}=2$ superconformal many--body mechanics
proceeds along different lines. It is relatively straightforward to extend the preceding analysis to
the case of a negative cosmological constant. However, in a spacetime with a positive cosmological constant
a conventional supersymmetric extension turns out to be
problematic \cite{gao}. The problem connects to the difficulty to define conserved
positive energy in the parent de Sitter space.
In the latter case we are led to consider a modified superalgebra in which anticommutator of
two supersymmetry generators yields the time translation generator
plus the generators of special conformal transformations and the $U(1)$--transformations.
This issue will be discussed in more detail in Sect. 7.

\vspace{0.5cm}

\noindent
5.1 {\it  Negative cosmological constant}\\

\noindent

The action functional of $\mathcal{N}=2$ mechanics in a spacetime with a negative cosmological constant is derived from (\ref{act3})
by adding oscillator potentials both for the bosonic and fermionic degrees of freedom
\bea\label{act4}
&&
S=\frac 12 \int dt \Big(\dot x^\a_i \dot x^\a_i-i \dot{\bar\p}^\a_i \p^\a_i+i\bar\p^\a_i {\dot\p}^\a_i
-\partial_{\a i} U(x) \partial_{\a i} U(x)-\frac{1}{R^2} x^\a_i x^\a_i+
\nonumber\\[2pt]
&& \qquad \qquad \qquad
+2 \partial_{\a i} \partial_{\b j} U(x) \p^\a_i \bar\p^\b_j
+\frac{2}{R} \p^\a_i \bp^\a_i  \Big).
\eea
As above, it is assumed that the prepotential $U(x)$ obeys the restrictions (\ref{Str2}).

It is straightforward to verify that
this action is invariant under transformations from the conformal Newton--Hooke
group\footnote{The fermionic degrees of freedom transform as vectors under space rotations.
They are intact under all other transformations from the
conformal Newton--Hooke group.} realized
as in Sect. 3.1. Also the $U(1)$--transformation (\ref{u1}) maintains its form.
As for symmetries with fermionic parameters, the supersymmetry transformation is slightly modified
\bea
&&
\d \p^\a_i=\Big(\dot x^\a_i -i\partial_{\a i} U(x)-\frac{i}{R} x^\a_i \Big) \bar\e, \qquad
\d \bar\p^\a_i=\Big(\dot x^\a_i +i\partial_{\a i} U(x) +\frac{i}{R} x^\a_i \Big) \e,
\nonumber\\[2pt]
&&
\d x^\a_i =i\Big(\p^\a_i \e+\bar\p^\a_i \bar\e\Big).
\eea
The superconformal transformation acquires the form
\bea
&&
\d \p^\a_i=\frac 12 R \Big(e^{-2it/R}-1\Big)\Big(\dot x^\a_i-i\partial_{\a i} U(x) -\frac{i}{R} x^\a_i\Big) \bar\kappa+i e^{-2it/R} x^\a_i \bar\kappa,
\nonumber\\[2pt]
&&
\d \bar\p^\a_i=\frac 12 R \Big(e^{2it/R}-1\Big)\Big(\dot x^\a_i+i\partial_{\a i} U(x) +\frac{i}{R} x^\a_i\Big) \kappa-i e^{2it/R} x^\a_i \kappa,
\nonumber\\[2pt]
&&
\d x^\a_i=\frac i2 R \Big(e^{2it/R}-1\Big)\p^\a_i \kappa+ \frac i2 R \Big(e^{-2it/R}-1\Big)\bar\p^\a_i \bar\kappa,
\eea
while translation in the fermionic sector now reads
\be
\d \p^\a_i=e^{-i t/R} \x^\a, \qquad \d \bar\p^\a_i=e^{i t/R}\bar\x^\a.
\ee

Below in Sect. 7 we shall be concerned with structure relations of the $\mathcal{N}=2$ superconformal
Newton--Hooke algebra. The easiest way to establish them is to work in a specific representation.
To this end, we display the Noether charges corresponding to the global symmetries given above
\bea\label{nc}
&&
H=\frac 12 \dot x^\a_i \dot x^\a_i+\frac 12 \partial_{\a i} U(x) \partial_{\a i} U(x)+\frac{1}{2 R^2} x^\a_i x^\a_i
-\partial_{\a i} \partial_{\b j} U(x) \p^\a_i \bar\p^\b_j-\frac{1}{R} \p^\a_i \bp^\a_i,
\nonumber\\[2pt]
&&
P^\a=\Big(\sum_{i=1}^n \dot x^\a_i \Big) \cos{(t/R)}+
\frac{1}{R} \Big( \sum_{i=1}^n x^\a_i \Big) \sin{(t/R)},
\nonumber\\[2pt]
&&
K^\a=\Big( \sum_{i=1}^n x^\a_i \Big) \cos{(t/R)}-\Big(\sum_{i=1}^n \dot x^\a_i \Big) R \sin{(t/R)},
\nonumber\\[2pt]
&&
M^{\a\b}=x^\a_i \dot x^\b_i-x^\b_i \dot x^\a_i-i\Big(\p^\a_i \bar\p^\b_i-\p^\b_i \bar\p^\a_i \Big),
\nonumber\\[2pt]
&&
D=-\frac 12 \Big( x^\a_i \dot x^\a_i \Big) \cos{(2t/R)} \ +\frac 12 \Big(H+\frac 1R J \Big) R \sin{(2t/R)}-\frac{1}{2 R} \Big( x^\a_i x^\a_i \Big) \sin{(2t/R)},
\nonumber\\[2pt]
&&
C=\frac 12 \Big( x^\a_i x^\a_i \Big) \cos{(2t/R)}-\frac 12 \Big( x^\a_i \dot x^\a_i \Big) R \sin{(2t/R)} -\frac 12 \Big(H+\frac 1R J \Big) R^2 (\cos{(2t/R)}-1),
\nonumber\\[2pt]
&&
Q=\Big(\dot x^\a_i +i \partial_{\a i} U(x)+\frac{i}{R} x^\a_i \Big) \p^\a_i, \qquad \quad ~
\bar Q=\Big(\dot x^\a_i -i \partial_{\a i} U(x)-\frac{i}{R} x^\a_i \Big) \bar\p^\a_i,
\nonumber\\[2pt]
&&
S=\Big(x^\a_i \p^\a_i \Big) e^{2it/R}+\frac{i}{2} Q R \Big(e^{2it/R}-1 \Big), \quad \bar S=\Big(x^\a_i \bar \p^\a_i \Big) e^{-2it/R}-
\frac{i}{2} \bar Q  R \Big(e^{-2it/R}-1 \Big),
\nonumber\\[2pt]
&&
L^\a=\Big(\sum_{i=1}^n \p^\a_i \Big) e^{it/R}, \quad \bar L^\a=\Big( \sum_{i=1}^n \bar\p^\a_i \Big) e^{-it/R}, \quad J=\p^\a_i \bp^\a_i.
\eea
Note that in the limit $R \to \infty$ both the symmetries and the
conserved charges fit to those of the $\mathcal{N}=2$ mechanics in a flat space, as they should.

\vspace{0.5cm}
\newpage
\noindent
5.2 {\it  Positive cosmological constant}\\

\noindent

The action functional of $\mathcal{N}=2$ mechanics in a spacetime with a positive cosmological constant is obtained from
(\ref{act4}) by reversing the sign of the oscillator potentials
\bea\label{act5}
&&
S=\frac 12 \int dt \Big(\dot x^\a_i \dot x^\a_i-i \dot{\bar\p}^\a_i \p^\a_i+i\bar\p^\a_i {\dot\p}^\a_i
-\partial_{\a i} U(x) \partial_{\a i} U(x)+\frac{1}{R^2} x^\a_i x^\a_i+
\nonumber\\[2pt]
&& \qquad \qquad \qquad
+2 \partial_{\a i} \partial_{\b j} U(x) \p^\a_i \bar\p^\b_j
-\frac{2}{R} \p^\a_i \bp^\a_i  \Big).
\eea
It holds invariant under the transformations from the conformal
Newton--Hooke group\footnote{Fermions are inert under all these transformations, but for rotations.} considered in Sect. 3.2.
The $U(1)$--symmetry acting on fermions maintains its form (\ref{u1}).

Among transformations involving odd parameters one finds the supersymmetry
\bea
&&
\d x^\a_i =i \p^\a_i  \Big(\cosh{(t/R)}-i \sinh{(t/R)} \Big)e^{-it/R} \e +i \bar\p^\a_i  \Big(\cosh{(t/R)}+i \sinh{(t/R)} \Big)e^{it/R} \bar\e,
\nonumber\\[2pt]
&&
\d \p^\a_i=\Big(\dot x^\a_i -i\partial_{\a i} U(x)\Big) \Big(\cosh{(t/R)}+i \sinh{(t/R)} \Big)e^{it/R}  \bar\e -
\nonumber\\[2pt]
&& \qquad \quad
-\frac iR x^\a_i \Big(\cosh{(t/R)}-i \sinh{(t/R)} \Big)e^{it/R} \bar\e ,
\nonumber\\[2pt]
&&
\d \bar\p^\a_i=\Big(\dot x^\a_i +i\partial_{\a i} U(x)\Big) \Big(\cosh{(t/R)}-i \sinh{(t/R)} \Big)e^{-it/R}  \e +
\nonumber\\[2pt]
&& \qquad \quad
+\frac iR x^\a_i \Big(\cosh{(t/R)}+i \sinh{(t/R)} \Big)e^{-it/R} \e ,
\eea
the superconformal transformations
\bea
&&
\d x^\a_i=-\p^\a_i R \sinh{(t/R)} e^{-it/R} \kappa+\bp^\a_i R \sinh{(t/R)} e^{it/R} \bar\kappa,
\nonumber\\[2pt]
&&
\d \p^\a_i=-i \Big(\dot x^\a_i-i\partial_{\a i} U(x) \Big)  R \sinh{(t/R)} e^{it/R} \bar\kappa
+i x^\a_i  \cosh{(t/R)} e^{it/R} \bar\kappa,
\nonumber\\[2pt]
&&
\d \bar\p^\a_i=i \Big(\dot x^\a_i+i\partial_{\a i} U(x) \Big)  R \sinh{(t/R)} e^{-it/R} \kappa
-i x^\a_i  \cosh{(t/R)} e^{-it/R} \kappa,
\eea
and the translations
\be
\d \p^\a_i=e^{i t/R} \x^\a, \qquad \d \bar\p^\a_i=e^{-i t/R}\bar\x^\a.
\ee

It is instructive to give the explicit form of the corresponding Noether charges
\bea\label{nc1}
&&
H=\frac 12 \dot x^\a_i \dot x^\a_i+\frac 12 \partial_{\a i} U(x) \partial_{\a i} U(x)-\frac{1}{2 R^2} x^\a_i x^\a_i
-\partial_{\a i} \partial_{\b j} U(x) \p^\a_i \bar\p^\b_j+\frac{1}{R} \p^\a_i \bp^\a_i,
\nonumber\\[2pt]
&&
P^\a=\Big(\sum_{i=1}^n \dot x^\a_i \Big) \cosh{(t/R)}-
\frac{1}{R} \Big( \sum_{i=1}^n x^\a_i \Big) \sinh{(t/R)},
\nonumber\\[2pt]
&&
K^\a=\Big( \sum_{i=1}^n x^\a_i \Big) \cosh{(t/R)}-\Big(\sum_{i=1}^n \dot x^\a_i \Big) R \sinh{(t/R)},
\nonumber\\[2pt]
&&
M^{\a\b}=x^\a_i \dot x^\b_i-x^\b_i \dot x^\a_i-i\Big(\p^\a_i \bar\p^\b_i-\p^\b_i \bar\p^\a_i \Big),
\nonumber
\eea
\bea
&&
D=-\frac 12 \Big( x^\a_i \dot x^\a_i \Big) \cosh{(2t/R)} +\frac 12 \Big(H-\frac 1R J \Big) R \sinh{(2t/R)}+\frac{1}{2 R} \Big( x^\a_i x^\a_i \Big) \sinh{(2t/R)},
\nonumber\\[2pt]
&&
C=\frac 12 \Big( x^\a_i x^\a_i \Big) \cosh{(2t/R)}-\frac 12 \Big( x^\a_i \dot x^\a_i \Big) R \sinh{(2t/R)} +\frac 12 \Big(H-\frac 1R J \Big) R^2 (\cosh{(2t/R)}-1).
\nonumber\\[2pt]
&&
Q=\Big(\dot x^\a_i +i \partial_{\a i} U(x)\Big) \p^\a_i \Big(\cosh{(t/R)}-i \sinh{(t/R)} \Big)e^{-it/R}+
\nonumber\\[2pt]
&& \qquad \quad +
i \p^\a_i x^\a_i \frac 1R
\Big(\cosh{(t/R)}+i \sinh{(t/R)} \Big)e^{-it/R},
\nonumber\\[2pt]
&&
\bar Q=
\Big(\dot x^\a_i -i \partial_{\a i} U(x)\Big) \bp^\a_i \Big(\cosh{(t/R)}+i \sinh{(t/R)} \Big)e^{it/R}-
\nonumber\\[2pt]
&& \qquad \quad -
i \bp^\a_i x^\a_i \frac 1R
\Big(\cosh{(t/R)}-i \sinh{(t/R)} \Big)e^{it/R},
\nonumber\\[2pt]
&&
S=\Big(x^\a_i \p^\a_i \Big) \cosh{(t/R)} e^{-it/R}-\Big(\dot x^\a_i +i \partial_{\a i} U(x)\Big) \p^\a_i R \sinh{(t/R)} e^{-it/R} ,
\nonumber\\[2pt]
&&
\bar S=\Big(x^\a_i \bp^\a_i \Big) \cosh{(t/R)} e^{it/R}-\Big(\dot x^\a_i -i \partial_{\a i} U(x)\Big) \bp^\a_i R \sinh{(t/R)} e^{it/R} ,
\nonumber\\[2pt]
&&
L^\a=\Big(\sum_{i=1}^n \p^\a_i \Big) e^{-it/R}, \quad \bar L^\a=\Big( \sum_{i=1}^n \bar\p^\a_i \Big) e^{it/R}, \quad J=\p^\a_i \bp^\a_i.
\eea
These will be used below in Sect. 7 when fixing structure relations of the $\mathcal{N}=2$ superconformal
Newton--Hooke algebra for the case of a positive cosmological constant.
Like their cousins (\ref{nc}), these functions transit smoothly to the generators of the $\mathcal{N}=2$ Schr\"odinger superalgebra (\ref{ch1})
in the limit $R \to \infty$.

\vspace{0.5cm}

\noindent
{\bf 6. Selected $\mathcal{N}=2$ prepotentials }\\

\noindent

As we have seen above, $\mathcal{N}=2$ superconformal many--body mechanics in Newton--Hooke spacetime is fully determined by a single
prepotential obeying the restrictions (\ref{Str2}). In this Section we discuss a special class of solutions to (\ref{Str2}) which are
related to root vectors of simple Lie algebras. We also extend the analysis in \cite{gl} and construct novel solutions which, after quantization,
generate  $\mathcal{N}=2$ models with only boson-fermion couplings.

Let us first discuss the rightmost equation in (\ref{Str2}). The general solution is a sum of a particular solution to the inhomogeneous equation
and the general solution to the homogeneous equation. The latter is given by an arbitrary function of the coordinate ratios $(x^\a_i/x^\b_j)$.
In this work, we narrow this wide class down to prepotentials of the special form
\be\label{35}
U(x)=\frac 12 \sum_\a g_\a \ln{(\a x,\a x)}.
\ee
In (\ref{35}) the sum runs over a set of vectors $\{\a_i \}$, $g_\a$ are coupling constants, and $(\a x,\a x)$ is a shorthand for
\be
\sum_{\b=1}^d \sum_{i,j=1}^n (\a_i x^\b_i) (\a_j x^\b_j).
\ee
Constraints (\ref{Str2}) then yield
\be\label{37}
\sum_{i=1}^n \a_i=0, \qquad \sum_\a g_\a=Z.
\ee
If one is concerned with systems of identical particles, then in addition to (\ref{37}) the set $\{\a \}$ is to be permutation invariant and
the couplings are to be identified in a proper way.

Interesting patterns of this kind are provided by root vectors of some simple Lie algebras.
First consider $A_{(n-1)}$ with $\{ \a\}=\{(1,-1,0,\dots,0)$ (plus permutations)$\}$ and couplings being all equal. The prepotential
\be
U(x)=\frac{g}{2} \sum_{i<j}^n\ln( x_{ij},x_{ij}),
\ee
where $x_{ij}^\a=x^\a_i-x^\a_j$, yields the bosonic potential in accord with (\ref{VB})
\be
V_{A_{(n-1)}} =g^2 \left( \sum_{i<j}^n \frac{1}{(x_{ij},x_{ij})}+\sum_{i\ne j ,i\ne k, j< k}^n
\frac{(x_{ij}, x_{i k})}{(x_{ij},x_{ij})(x_{i k},x_{ik})}\right).
\ee
This is the model of Calogero and Marchioro \cite{cm} in which one
disregards the pairwise harmonic interaction and identifies the couplings which control the two--body and three--body
interactions\footnote{The identification of couplings is typical for superconformal mechanics (see e.g. the discussion in \cite{gala2}).}.

Our second example is a three--body model based on $\{ \a \}=\{(1,-1,0)$ (plus permutations)$\}$ and  $\{ \b\}=\{(1,1,-2)$ (plus permutations)$\}$
which are the short and the long root vectors of $G_2$. In this case two independent couplings may enter the prepotential
\be
U(x)=\frac{1}{2} \Big( g_S \sum_{i<j}^3\ln( x_{ij},x_{ij})+g_L \sum_{k\ne i ,k\ne j, i< j}^3 \ln( x_{ijk},x_{ijk})   \Big),
\ee
where $x^\a_{ijk}=x^\a_i+x^\a_j-2 x^\a_k$, which gives rise to the bosonic potential
\bea
&&
V_{G_2} =g_S^2 \left( \sum_{i<j}^3 \frac{1}{(x_{ij},x_{ij})}+\sum_{i\ne j ,i\ne k, j< k}^3 \frac{(x_{ij}, x_{i k})}{(x_{ij},x_{ij})(x_{i k},x_{ik})}\right)
+3 g_S g_L \sum_{k\ne i \ne j}^3 \frac{(x_{kij}, x_{ij})}
{(x_{kij},x_{kij})(x_{ij},x_{ij})}
\nonumber\\[2pt]
&& \qquad \quad
+3 g_L^2 \left( \sum_{k\ne i ,k\ne j, i< j}^3 \frac{1}{(x_{ijk},x_{ijk})}-\sum_{i\ne j ,i\ne k, j< k}^3 \frac{(x_{ijk}, x_{i kj})}
{(x_{ijk},x_{ijk}) (x_{i kj},x_{ikj})}\right).
\eea
This can be viewed as a multidimensional generalization of the model studied by Wolfes in \cite{W}.

In some instances in order to conform to the leftmost constraint in (\ref{37}), one has to extract a reasonable subset from the full set of root vectors of a
simple Lie algebra.
For example, in the case of $F_4$ one can take $\{ \a \}=\{(1,-1,0,0)$ (plus permutations)$\}$ and  $\{ \b\}=\{(1,1,-1,-1)$ (plus permutations)$\}$
and form the prepotential
\be
U(x)=\frac{1}{2} \Big( g_S \sum_{i<j}^4\ln( x_{ij},x_{ij})+g_L \sum_{i<j, i<k, k<l}^4 \ln( x_{ijkl},x_{ijkl})   \Big),
\ee
where $x^\a_{ijkl}=x^\a_i+x^\a_j-x^\a_k-x^\a_l$. This yields the bosonic potential
\bea
&&
V_{F_4} =g^2_S \left( \sum_{i<j}^4 \frac{1}{(x_{ij},x_{ij})}+\sum_{i\ne j ,i\ne k, j< k}^4 \frac{(x_{ij}, x_{i k})}{(x_{ij},x_{ij})(x_{i k},x_{ik})}\right)
+2g_L^2 \sum_{i<j,k<l,i\ne k}^4 \frac{1}{(x_{ijkl},x_{ijkl})}+
\nonumber\\[2pt]
&& \qquad \quad +
2 g_S g_L \sum_{k\ne l \ne i\ne j, i<j}^4 \frac{(x_{ij},x_{iklj})}{(x_{ij},x_{ij})(x_{iklj},x_{iklj})},
\eea
which is a generalization of Wolfes' four--body model \cite{W1} to arbitrary dimension.

When constructing quantum theory,  $\mathcal{N}=2$ fermions $\p^\a_i$, $\bp^\a_i$ are conventionally realized as
creation and annihilation operators acting in the Fock space. Symmetric ordering in the sector of fermionic variables
then yields a quantum correction to the classical bosonic potential \cite{gl} (see also Sect. 7)
 \be
V_B=\frac 12 (\partial_{\a i} U(x) \partial_{\a i} U(x)+\hbar \partial_{\a i} \partial_{\a i} U(x)).
\ee

A special class of $\mathcal{N}=2$ models arises if one takes the prepotential in the form
\be
U(x)=\hbar \ln{G(x)},
\ee
where the function $G(x)$ obeys Laplace's equation and the homogeneity condition
\be\label{G}
\Delta G(x)=0, \qquad x_i^\a \partial_{\a i} G(x)=\frac{Z}{\hbar} G(x).
\ee
It is assumed that $G(x)$ is invariant under translations and rotations. Obviously,
in this case $V_B=0$ and the resulting system has only boson--fermion couplings. Furthermore,
choosing $G(x)$ to be a homogeneous polynomial of degree $l$, i.e. $x_i^\a \partial_{\a i} G(x)=l G(x)$,
one arrives at a class of models for which $Z$ is naturally quantized  $Z=\hbar l$.

In Ref. \cite{gl} particular solutions to (\ref{G}) were constructed. We conclude this Section by providing
two new examples of this kind which are linked to $G_2$ and $F_4$. Our starting point is the homogeneous polynomial of degree
$l$ which involves two sets of vectors $\{ \a\}$ and  $\{ \b\}$
\be\label{g2}
G(x)=\sum_\b (\b x,\b x)^{\frac{l}{2}} +\lambda \sum_\a (\a x,\a x)^{\frac{l}{2}}.
\ee
Here $\lambda$ is a real constant to be fixed below.
Consider first the short and the long positive root vectors of $G_2$:
$\{\a\}=\{(1,-1,0),(1,0,-1),(0,1,-1) \}$, $\{ \b\}=\{(1,1,-2),(1,-2,1),(2,-1,\\
-1)\}$. Taking into account the identity
\be
\sum_\b (\b x,\b x)^2=9 \sum_\a (\a x,\a x)^2,
\ee
one can readily verify that the ansatz (\ref{g2}) solves Laplace's equation provided
\be
l=6, \qquad \lambda=-27.
\ee
Note that this choice of the parameters does yield a nontrivial three--body model with a non--vanishing
$G(x)$. One could try to construct a similar
solution using a polynomial of the fourth order and the identity
$\sum_\b (\b x,\b x)=3 \sum_\a (\a x,\a x)$. However, the latter leads to the trivial solution $G(x)=0$.

Our last example is based on the subset of the positive root vectors of $F_4$:  $\{\a\}=\{(1,-1,0,0),
(1,0,-1,0),(1,0,0,-1),(0,1,-1,0),(0,1,0,-1),(0,0,1,-1) \}$ and $\{ \b\}=\\
\{(1,1,-1,-1),(1,-1,1,-1),(1,-1,-1,1)\}$.  In this case the key identity involving the short and long root vectors
reads
\be
\sum_\b (\b x,\b x)=\sum_\a (\a x,\a x).
\ee
It prompts us to try the ansatz (\ref{g2}) with $l=4$. Laplace's equation then determines $\lambda=-2$.

\vspace{0.5cm}

\noindent
{\bf 7. $\mathcal{N}=2$ superconformal Newton--Hooke algebra}\\

\noindent

In this Section we establish structure relations of the $\mathcal{N}=2$ superconformal Newton--Hooke algebra.
To this end, we consider quantum mechanical representation of the conserved charges given in Sect. 5 and calculate
their (anti)commutators. Constraints (\ref{Str2}) on the prepotential should be remembered.
For the bosonic operators the standard coordinate representation is chosen
\be\label{cr}
[x^\a_i,p^\b_j]=i\hbar \d^{\a\b} \d_{ij},
\ee
with $p^\a_i=-i\hbar\partial_{\a i}$. $\mathcal{N}=2$ fermions are realized as
creation--annihilation operators acting in the Fock space\footnote{When analyzing the models (\ref{act4}) and (\ref{act5}) within
the Hamiltonian formalism, one finds fermionic second class constraints. Introducing the Dirac bracket and resolving the constraints one can
remove
momenta canonically conjugate to the configuration space variables $\p^\a_i$ and $\bp^\a_i$. This leads one to the bracket (\ref{pp}).}
\be\label{pp}
\{\p^\a_i,\bp^\b_j \}=\hbar \d^{\a\b} \d_{ij},
\ee
with ${\Big(\p^\a_i \Big)}^{\dagger}= \bp^\a_i$.
In order to get a closed algebra after quantization one has to use the Weyl ordering.
Given the functions (\ref{nc}) and (\ref{nc1}), this implies
\bea
&&
x^\a_i \dot x^\a_i \quad \rightarrow \quad \frac 12 \Big(x^\a_i p^\a_i+p^\a_i x^\a_i \Big),
\nonumber\\[2pt]
&&
\p^\a_i \bp^\b_j \quad \rightarrow \quad \frac 12 \Big(\p^\a_i \bp^\b_j-\bp^\b_j \p^\a_i \Big),
\eea
where the rightmost expressions involve quantum operators obeying the (anti)commutation relations (\ref{cr}) and (\ref{pp}).
Now we are in a position to compute structure relations of the
 $\mathcal{N}=2$ superconformal Newton--Hooke algebra.

\vspace{0.5cm}

\noindent
7.1 {\it  Negative cosmological constant}\\

\noindent

In order to avoid the appearance of a fictitious central charge in structure relations of the superalgebra, we shift
the Hamiltonian $H$ and the $U(1)$ generator $J$ considered in Sect. 5.1 by a constant $Z$ from (\ref{Str2})
\be
H \quad \rightarrow \quad H+\frac{Z}{R}, \qquad J \quad \rightarrow \quad J-Z,
\ee
such that the combination $H+\frac{1}{R} J$ entering $D$ and $C$ is intact. Then
a straightforward calculation yields (vanishing (anti)commutators are omitted)
\begin{align}\label{algebra1}
&
\{Q,\bar Q \}=2\hbar H, && \{Q,\bar S \}=-2\hbar D-i\hbar J+ \frac{2i\hbar}{R} C,
\nonumber\\[2pt]
&
\{S,\bar S \}=2\hbar C, && \{\bar Q,S \}=-2\hbar D+i\hbar J- \frac{2i\hbar}{R} C,
\nonumber\\[2pt]
&
\{Q,\bar L^\a\}=\hbar (P^\a+\frac iR  K^\a ), && \{\bar Q,L^\a\}=\hbar (P^\a-\frac iR  K^\a ),
\nonumber\\[2pt]
&
\{ S,\bar L^\a \}=\hbar K^\a, && \{\bar S,L^\a \}=\hbar K^\a,
\nonumber\\[2pt]
&
\{L^\a, \bar L^\b \}=\hbar \delta^{\a\b} M_1, && [P^\a,K^\b]=-i\hbar \d^{\a\b} M,
\nonumber\\[2pt]
&
[D,Q]=-\frac{i\hbar}{2} Q-\frac{\hbar}{R} S, && [D,\bar Q]=-\frac{i\hbar}{2} \bar Q+\frac{\hbar}{R} \bar S,
\nonumber\\[2pt]
&
[C,Q]=i\hbar S, && [C,\bar Q]=i\hbar \bar S,
\nonumber\\[2pt]
&
[J,Q]=\hbar Q, && [J,\bar Q]=-\hbar \bar Q,
\nonumber\\[2pt]
&
[Q,P^\a]=-\frac{\hbar}{R} L^\a, && [\bar Q,P^\a]=\frac{\hbar}{R} \bar L^\a,
\nonumber\\[2pt]
&
[Q,K^\a]=-i \hbar L^\a, && [\bar Q,K^\a]=-i \hbar \bar L^\a,
\nonumber\\[2pt]
&
[D,S]=\frac{i\hbar}{2} S, && [D,\bar S]=\frac{i\hbar}{2} \bar S,
\nonumber\\[2pt]
&
[H,S]=-i\hbar (Q-\frac{2i}{R} S), && [H,\bar S]=-i\hbar (\bar Q+\frac{2i}{R} \bar S),
\nonumber\\[2pt]
&
[J,S]=\hbar S, && [J,\bar S]=-\hbar \bar S,
\nonumber\\[2pt]
&
[S,P^\a]=i \hbar L^\a, && [\bar S,P^\a]=i \hbar \bar L^\a,
\nonumber\\[2pt]
&
[J,L^\a]=\hbar L^\a, && [J,\bar L^\a]=-\hbar \bar L^\a,
\nonumber\\[2pt]
&
[H,L^\a]=-\frac{\hbar}{R} L^\a, && [H,\bar L^\a]=\frac{\hbar}{R} \bar L^\a,
\nonumber
\end{align}
\begin{align}
&
[D,P^\a]=-\frac{i\hbar}{2} P^\a, && [C,P^\a]=i\hbar K^\a,
\nonumber\\[2pt]
&
[H,P^\a]=\frac{i\hbar}{R^2} K^\a, && [H,K^\a]=-i\hbar P^\a,
\nonumber\\[2pt]
&
[D,K^\a]=\frac{i\hbar}{2} K^\a, && [H,D]=i\hbar (H+\frac 1R J-\frac{2}{R^2} C),
\nonumber\\[2pt]
&
[H,C]=2i\hbar D, && [D,C]=i\hbar C,
\nonumber\\[2pt]
&
[M^{\a\b},P^\g]=i\hbar(\d^{\a\g} P^\b-\d^{\b\g} P^\a), && [M^{\a\b},K^\g]=i\hbar(\d^{\a\g} K^\b-\d^{\b\g} K^\a),
\nonumber\\[2pt]
&
[M^{\a\b},L^\g]=i\hbar(\d^{\a\g} L^\b-\d^{\b\g} L^\a), && [M^{\a\b},\bar L^\g]=i\hbar(\d^{\a\g} \bar L^\b-\d^{\b\g} \bar L^\a),
\nonumber\\[2pt]
&
[M^{\a\b},M^{\g\d}]=i\hbar(\d^{\a\g} M^{\b\d}+\d^{\b\d} M^{\a\g}-
\nonumber\\[2pt]
&
\qquad \qquad \qquad -\d^{\b\g} M^{\a\d}-\d^{\a\d} M^{\b\g}).
\end{align}
Above $M$ and $M_1$ are central charges\footnote{For the particular representation considered in Sect. 5.1 the central charges read $M=M_1=n$.}.
This superalgebra differs from the $N=2$ Schr\"odinger
superalgebra \cite{gm} only by contributions which explicitly involve the factors $\frac{1}{R}$ and $\frac{1}{R^2}$.
In the flat space limit $R \to \infty$ they disappear and the two superalgebras merge, as they should.

\vspace{0.5cm}

\noindent
7.2 {\it  Positive cosmological constant}\\

\noindent

For the case of a positive cosmological constant we also shift
the Hamiltonian and the $U(1)$ generator defined in Sect. 5.2 by a constant $Z$
\be
H \quad \rightarrow \quad H-\frac{Z}{R}, \qquad J \quad \rightarrow \quad J-Z.
\ee
Note that  $H-\frac{1}{R} J$ which appears in $D$ and $C$ does not feel the change. Then
we take into account the constraints (\ref{Str2}) and calculate the structure relations
\begin{align}\label{algebra2}
&
\{Q,\bar Q \}=2\hbar (H-\frac{2}{R} J+\frac{2}{R^2} C), && \{Q,\bar S \}=-2\hbar D-i\hbar J+ \frac{2i\hbar}{R} C,
\nonumber\\[2pt]
&
\{S,\bar S \}=2\hbar C, && \{\bar Q,S \}=-2\hbar D+i\hbar J- \frac{2i\hbar}{R} C,
\nonumber\\[2pt]
&
\{Q,\bar L^\a\}=\hbar (P^\a+\frac iR  K^\a ), && \{\bar Q,L^\a\}=\hbar (P^\a-\frac iR  K^\a ),
\nonumber\\[2pt]
&
\{ S,\bar L^\a \}=\hbar K^\a, && \{\bar S,L^\a \}=\hbar K^\a,
\nonumber\\[2pt]
&
\{L^\a, \bar L^\b \}=\hbar \delta^{\a\b} M_1, && [P^\a,K^\b]=-i\hbar \d^{\a\b} M,
\nonumber\\[2pt]
&
[H,Q]=\frac{2\hbar}{R} (Q-\frac{i}{R} S), && [H,\bar Q]=-\frac{2\hbar}{R} (\bar Q+\frac{i}{R}\bar S),
\nonumber\\[2pt]
&
[D,Q]=-\frac{i\hbar}{2} Q-\frac{\hbar}{R} S, && [D,\bar Q]=-\frac{i\hbar}{2} \bar Q+\frac{\hbar}{R} \bar S,
\nonumber\\[2pt]
&
[C,Q]=i\hbar S, && [C,\bar Q]=i\hbar \bar S,
\nonumber
\end{align}
\begin{align}
&
[J,Q]=\hbar Q, && [J,\bar Q]=-\hbar \bar Q,
\nonumber\\[2pt]
&
[Q,P^\a]=-\frac{\hbar}{R} L^\a, && [\bar Q,P^\a]=\frac{\hbar}{R} \bar L^\a,
\nonumber\\[2pt]
&
[Q,K^\a]=-i \hbar L^\a, && [\bar Q,K^\a]=-i \hbar \bar L^\a,
\nonumber\\[2pt]
&
[D,S]=\frac{i\hbar}{2} S, && [D,\bar S]=\frac{i\hbar}{2} \bar S,
\nonumber\\[2pt]
&
[H,S]=-i\hbar Q, && [H,\bar S]=-i\hbar \bar Q,
\nonumber\\[2pt]
&
[J,S]=\hbar S, && [J,\bar S]=-\hbar \bar S,
\nonumber\\[2pt]
&
[S,P^\a]=i \hbar L^\a, && [\bar S,P^\a]=i \hbar \bar L^\a,
\nonumber\\[2pt]
&
[J,L^\a]=\hbar L^\a, && [J,\bar L^\a]=-\hbar \bar L^\a,
\nonumber\\[2pt]
&
[H,L^\a]=\frac{\hbar}{R} L^\a, && [H,\bar L^\a]=-\frac{\hbar}{R} \bar L^\a,
\nonumber\\[2pt]
&
[D,P^\a]=-\frac{i\hbar}{2} P^\a, && [C,P^\a]=i\hbar K^\a,
\nonumber\\[2pt]
&
[H,P^\a]=-\frac{i\hbar}{R^2} K^\a, && [H,K^\a]=-i\hbar P^\a,
\nonumber\\[2pt]
&
[D,K^\a]=\frac{i\hbar}{2} K^\a, && [H,D]=i\hbar (H-\frac 1R J+\frac{2}{R^2} C),
\nonumber\\[2pt]
&
[H,C]=2i\hbar D, && [D,C]=i\hbar C,
\nonumber\\[2pt]
&
[M^{\a\b},P^\g]=i\hbar(\d^{\a\g} P^\b-\d^{\b\g} P^\a), && [M^{\a\b},K^\g]=i\hbar(\d^{\a\g} K^\b-\d^{\b\g} K^\a),
\nonumber\\[2pt]
&
[M^{\a\b},L^\g]=i\hbar(\d^{\a\g} L^\b-\d^{\b\g} L^\a), && [M^{\a\b},\bar L^\g]=i\hbar(\d^{\a\g} \bar L^\b-\d^{\b\g} \bar L^\a),
\nonumber\\[2pt]
&
[M^{\a\b},M^{\g\d}]=i\hbar(\d^{\a\g} M^{\b\d}+\d^{\b\d} M^{\a\g}-
\nonumber\\[2pt]
&
\qquad \qquad \qquad -\d^{\b\g} M^{\a\d}-\d^{\a\d} M^{\b\g}),
\end{align}
where $M$ and $M_1$ are central charges\footnote{For the particular representation considered in Sect. 5.2 the central charges are $M=M_1=n$.}.

A few comments are in order. Firstly,
as compared to the previous case, only the (anti)com\-mu\-tators $\{Q,\bar Q \}$,
$[H,Q]$, $[H,S]$, $[H,L^\a]$, $[H,D]$, $[H,P^\a]$ and their Hermitian conjugates are altered. Most notable are the changes in
$\{Q,\bar Q \}$, $[H,Q]$ and $[H,\bar Q]$ which modify the standard $\mathcal{N}=2$ superalgebra. As was mentioned in the Introduction, this
modification
is necessary in order to reconcile the fact that the spectrum of the original Hamiltonian is not bounded from below,
while  $\{Q,\bar Q\}$ is a positive definite
operator. Note that the modification essentially uses the dimensionful constant $R$ which is available only in Newton--Hooke
spacetime. Secondly, the appearance of the special conformal generator $C$ on the right--hand side of $\{Q,\bar Q\}$ resembles
a modification of the Hamiltonian of a single conformal particle in one dimension proposed in \cite{ddf}. In that framework it was used to
cure the ground state problem of the conformal particle. Thirdly, in the flat space limit $R \to \infty$ the superalgebra
correctly reproduces the $N=2$ Schr\"odinger
superalgebra.

\vspace{0.5cm}

\noindent
{\bf 8. Decoupling transformation on $\mathcal{N}=2$ mechanics in Newton--Hooke spacetime}\\

\noindent

We conclude this work by briefly discussing a decoupling transformation
on $\mathcal{N}=2$ quantum mechanics in Newton--Hooke spacetime.
Such a transformation was originally proposed for the
Calogero model in one dimension \cite{glp1} (see also related works \cite{pol,gur}).
This model involves a conformal potential describing pairwise interaction of identical particles
and the harmonic potential \cite{calo}. A similarity transformation in \cite{glp1,gur} relates it
to a set of decoupled oscillators. It offers a simple and straightforward method of building the complete set of
eigenstates \cite{gur} and provides
an efficient means for constructing various $N=2$ and $N=4$ superconformal
many--body models in one dimension \cite{glp1,glp}.
An elegant geometric interpretation
of the similarity transformation as the inversion of the Klein model of the Lobachevsky
plane was proposed in \cite{arm,bn}.
It is amazing that the existence of such a transformation
was anticipated by Calogero in his original work \cite{calo}.

First let us treat the case of a negative cosmological constant.
A key ingredient of the construction is the conformal algebra $so(1,2)$
\bea
[H_0,C_0]=2i\hbar D_0, \qquad [H_0,D_0]=i\hbar H_0, \qquad [D_0,C_0]=i\hbar C_0,
\eea
and its representation on a set of decoupled particles
\bea
H_0=\frac 12 p^\a_i p^\a_i, \qquad D_0=-\frac 14 \Big(x^\a_i p^\a_i+p^\a_i x^\a_i \Big), \qquad C_0=\frac 12 x^\a_i x^\a_i.
\eea
As the first step one considers an automorphism of the $\mathcal{N}=2$ superconformal Newton--Hooke algebra
\be\label{ser}
T \quad \rightarrow \quad T'=e^{\frac{i}{\hbar}A}~ T~ e^{-\frac{i}{\hbar}A}=T+\sum_{n=1}^\infty\frac{1}{n!}{\left(\frac{i}{\hbar}\right)}^n
\underbrace{[A,[A, \dots [A,T] \dots]}_{n~\rm times},
\ee
generated by a specific linear combination of the operators $H$, $J$, $D_0$ and $C_0$
\be\label{a}
A=\frac{i R}{2}(H+\frac 1R J) -\frac{i}{R} C_0 -D_0.
\ee
Here $H$ and $J$ are realized as in Sect. 7.1. The number coefficients in (\ref{a}) are adjusted so as to terminate
the series in the Baker--Hausdorff formula at a final step\footnote{To be more precise, one takes $A$ to be a
linear combination  of $H$, $D_0$, $C_0$ and $J$ with arbitrary coefficients. Then one demands the series in (\ref{ser})
to terminate at most at the third step. This fixes two coefficients in terms of the others. Then one requires the generator
$H$ to disappear from $H'$ which determines one more coefficient. In general this procedure leaves one coefficients in the transformation
operator $A$ arbitrary. The particular choice (\ref{a}) is taken for calculational simplicity.}. In particular, for the Hamiltonian $H$ this gives
\be
H'=\frac{2}{R^2} C_0-\frac{2i}{R} D_0-\frac 1R J.
\ee

That the interacting Hamiltonian is mapped to the linear combination of $C_0$, $D_0$ and $J$ prompts one to
try a similar transformation involving $H_0$, $D_0$, $C_0$ and $J$
\be\label{sec}
T' \quad \rightarrow \quad T^{''}=e^{\frac{i}{\hbar}B}~ T'~ e^{-\frac{i}{\hbar}B}, \qquad B=-\frac{i R}{2} (H_0+\frac 1R J)+\frac{i}{2R}  C_0+D_0.
\ee
It turns out that such a choice of $B$ again terminates the infinite series at a final step and yields
\bea
&&
H''=H_0+\frac{1}{2 R^2} x^\a_i x^\a_i-\frac 1R J.
\eea
This is the original Hamiltonian $H$ in which the conformal potential is discarded.
Thus with the use of the operator $e^{\frac{i}{\hbar} B} e^{\frac{i}{\hbar} A}$ one can map
the Hamiltonian of a generic $\mathcal{N}=2$ mechanics in Newton--Hooke spacetime to that of
decoupled (super)oscillators. An immediate corollary is that the energy eigenvalues
are the same in both the pictures. Another corollary is that the original many--body model
is quantum integrable.

The case of a positive cosmological constant is treated along similar lines.
First one considers a similarity transformation generated by
\be\label{a1}
A=\frac{R}{2}(H-\frac 1R J) +\frac{1}{R} C_0 -D_0,
\ee
where $H$ and $J$ are defined as in Sect. 7.2. Being applied to the Hamiltonian, this yields
\be
H'=-\frac{2}{R^2} C_0+\frac 2R D_0+\frac 1R J.
\ee
Then the second map which involves
\be
B=-\frac{R}{2} (H_0-\frac 1R J)-\frac{1}{2R}  C_0+D_0
\ee
provides the desired transformation
\be
H''=H_0-\frac{1}{2 R^2} x^\a_i x^\a_i+\frac 1R J.
\ee

In this work we do not consider how other operators
which form a representation of the $\mathcal{N}=2$ superconformal Newton--Hooke algebra
transform under the similarity transformation. Suffice it to mention they will acquire nonlocal contributions
involving the conformal potential. Thus the simplification in the dynamics is achieved at
the price of a nonlocal realization of the full $\mathcal{N}=2$ superconformal Newton--Hooke algebra in a
Hilbert space.

\vspace{0.5cm}

\noindent
{\bf 9. Conclusion}\\

To summarize, in this work we have systematically studied conformal many--body
mechanics
in Newton--Hooke spacetime. Constraints imposed on the interaction potential by the conformal Newton--Hooke
symmetry were found. The Lagrangian, its global symmetries and the conserved
charges were given in a form convenient for analyzing the flat space limit.
$\mathcal{N}=2$ superconformal extension was discussed in detail. It was shown that
in the case of a positive cosmological constant one has to modify the conventional
superalgebra. Structure relations of the $\mathcal{N}=2$ superconformal
Newton--Hooke algebra were given in the form in which the flat space limit is unambiguous.
A new class of  $\mathcal{N}=2$ models in arbitrary dimension which is related to root vectors of
simple Lie algebras was presented. Automorphisms of the $\mathcal{N}=2$ superconformal Newton-Hooke
algebra generated by the conformal algebra so(1,2) were studied. A similarity transformation which
allows one to establish the structure of the energy spectrum was proposed.

Let us mention a few possible developments of this work. First of all, it is
interesting to generalize the present analysis to the case of spinning particles in
Newton--Hooke spacetime.
Then it is worth studying larger (super)conformal Newton--Hooke algebras, including
an infinite dimensional one, and their dynamical realizations.
As was demonstrated in \cite{NI,PER} the whole motion of a free particle
can be associated to a half-period of the harmonic oscillator via a specific local diffeomorphism.
With the use of this transformation one can construct
wave function of the oscillator starting from that of a free particle.
A possibility to generalize the transformation  in \cite{NI,PER}
to the case of many-body conformal mechanics in the harmonic trap is an interesting open problem.
Finally, it is tempting to construct an off-shell superfield formulation for
$\mathcal{N}=2$ mechanics in Newton--Hooke spacetime.

\vspace{0.5cm}

\noindent{\bf Acknowledgements}\\

\noindent
This work was supported in part by 
the Dynasty Foundation and RFBR grant 09-02-00078.

\end{document}